# The Infrastructure Equation:
# Water, Energy, and Community Policy for Georgia's Data Center Boom

Mickey M. Rogers, William M. Ota, Nathaniel Burola, Tepring Piquado

The rapid growth of data centers driven by cloud computing and artificial intelligence is reshaping infrastructure planning and environmental governance in the United States. Georgia has emerged as a major market for data center development, particularly in the Atlanta metropolitan region, creating economic opportunity alongside significant challenges. Data centers are water-intensive, energy-intensive, and land-intensive infrastructure whose cumulative impacts strain municipal water systems, electric grids, and local land-use frameworks. Unlike single industrial projects, data centers are often proposed in clusters, amplifying community and infrastructure impacts.

This report draws on insights from a Georgia-based expert convening to describe the implications of data center growth for water management, energy reliability, ratepayer equity, zoning, and community engagement, identify potential gaps in transparency and regulatory coordination, and present a policy roadmap to help Georgia balance digital infrastructure growth with sustainability, equity, and community protection.

**Corresponding author:** tepring@scitechpolicy.org

**Acknowledgements:** The authors thank Dr. Natalie Seitzman and Olivia Ascher for their thoughtful review and valuable feedback on this report. We are grateful to Taylor Spicer, Executive Director of Engineers and Scientists Acting Locally (ESAL), and Tepring Piquado, CEO of the National Science Policy Network (NSPN), for co-organizing the convening and guiding the project from concept to completion. We extend our appreciation to Amy Sharma, Executive Director at Science for Georgia, for her collaboration and contributions to the design, facilitation, and moderation of the workshop. Finally, we thank Jen Bonnet at Invest Atlanta for supporting the convening at the Atlanta Tech Hub, an environment that fostered productive dialogue and shared learning. We also acknowledge the Alfred P. Sloan Foundation, whose support made the event and subsequent analysis possible.

## I.  Introduction

Data centers matter because they quietly keep almost everything in our daily lives running. Every time someone pays a bill online, looks up a medical record, streams a movie, makes a payment, or asks an artificial intelligence (AI) tool a question, that activity depends on servers cranking out computations inside a data center somewhere. These facilities have become the digital backbone of modern life. Without them, schools would struggle to teach, hospitals would struggle to care for patients, and businesses of every size would struggle to operate.

Data centers are playing an important role in Georgia's economic future. A data center may not employ hundreds of people on-site, but the jobs it does create can be highly skilled and well paid, such as a facilities operations engineer or a data technical (Mayer & Velkova, 2023). Large construction projects like data centers brings in years of work for local trades (Erlich & Grabelsky, 2005) and their long-term presence has shown to attract other tech companies desiring proximity to reliable computing power and high-quality network infrastructure. When a region has strong digital capacity, it becomes appealing to industries that depend on fast, secure, and high-volume computing (Greenstein & Fang, 2022; Walcott & Wheeler, 2001).

To support new data centers, utilities may expand fiber networks, upgrade energy substations, or modernize water systems. When these upgrades are planned with communities in mind, they can strengthen the reliability of services for everyone, not just for a single project. This could be why states are competing for data center investment. If developed thoughtfully, these facilities can help grow local economies, support innovation, and give residents access to the digital tools and services that are becoming essential in everyday life.

The rapid rise in AI adoption and cloud computing services has significantly increased demand for data center infrastructure in the U.S. Spending on data center construction has soared in the past decade from $1.8 billion in 2014 to $28.3 billion in 2024 (Muschter, 2025). This expansion has raised policy concerns related to energy consumption, environmental impacts, and the capacity of supporting infrastructure as the U.S. seeks to maintain a competitive position in the global AI race.

Data centers are "large buildings that house rows of computer servers, data storage systems and networking equipment, as well as the power and cooling systems that keep them running" (Leppert, 2025). Despite heightened attention in recent years, data centers have existed since the early 1990s (Robertson, 2024). There are different types of data centers; some common types include (Leppert, 2025):

1. **Hyperscale data centers:** Warehouse-sized facilities that store advanced servers capable of handling massive processing workloads. This model has been growing in popularity with the emergence of cloud computing services, cryptocurrency mining, and AI workloads for the data center boom.
2. **Enterprise data centers:** Facilities that are owned and operated by businesses for their private storage and computing needs.
3. **Colocation data centers/service providers:** Facilities that are rented out to individual businesses.
4. **Edge data centers:** Compact, decentralized facilities that process data closer to end users, reducing latency and boosting speed for real-time applications and local computing needs.

The precise number of data centers in the U.S. can fluctuate. Estimates place the number between 4,165 and 4,214 as of November 2025 (Taylor, 2025; *USA Data Centers*, n.d.). A third of U.S. data centers are located in three states: Virginia, Texas, and California (Leppert, 2025). Companies choose data center locations based on a range of factors, including the availability of capable power utilities, properly zoned land, and high-quality network access to reduce latency.

Data centers represent critical infrastructure for the digital economy which is poised for significant growth. According to McKinsey & Company, companies will invest almost $7 trillion in global capital expenditures on data center infrastructure. More than $4 trillion will go towards computing-hardware investments focused on real estate and power infrastructure (Remington & Carter, 2025)2/10/2026 11:49:00 PM. States that plan, manage, and mitigate the risks of data center growth stand to profit from economic growth, the creation of high-paying jobs, and continued research and innovation.

The Commercial Real Estate Development Association (NAIOP) analyzed 36 states that have some form of legislation authorizing tax incentives for new data center development (Remington & Carter, 2025)2/10/2026 11:49:00 PM. There is no standard template for structuring incentives. However, this approach has gathered momentum as stakeholders seek to benefit from this data center boom. States will differ in how they address key issues including the kind of tax exemption offered, job creation thresholds, and the kinds of expenditures covered in state legislation for data center companies (*Tax Incentives for Data Centers 50 State Survey*, n.d.). Regardless, Jeff Jakubiak notes in Utility Dive that data centers are seen as engines of economic growth, "playing a role similar to what steel mills and manufacturing plants once did — anchoring regional economies and signaling innovation" (Jakubiak, 2025).

Despite these opportunities, there are significant challenges in scaling data center infrastructure. According to Data Center Watch, opposition to data center projects is accelerating with an "estimated $98 billion in projects [that] were blocked or delayed, more than the total for all previous quarters since 2023". More than 53 active groups across 17 states targeted 30 data center projects in Q2 of 2025 (*Data Center Watch Report Q2 2025*, 2025).

A key constraint for states is power availability. Epoch AI estimates that in the U.S., AI data centers will collectively need around 20-30 gigawatts of power by late 2027. To meet short-term energy demands, developers are building natural gas power plants on-site while they wait to be connected to the national power grid (Ho, 2025). To meet long-term energy demands, tech companies have invested significant capital in nuclear energy development. As an example, Microsoft has partnered with Constellation Energy to revive the Three Mile Island Nuclear Power Plant in Pennsylvania along with signing purchasing agreements with nuclear power startups (Mandler, 2024).

Data centers require a significant amount of water for cooling purposes to prevent overheating (Iftekhar & Browne, 2025). Densely packed servers generate substantial heat, requiring cooling systems that withdraw and consume large amounts of water. According to WestWater Research, data center water consumption in the U.S. is projected to increase by 170 percent between 2023 and 2030 (Iftekhar & Browne, 2025). Recent estimates suggest that accelerated AI adoption could result in an additional 4.2 – 6.6 billion cubic meters of water withdrawal by 2027, including onsite cooling and offsite electricity generation (Spindler et al., 2025).

There is a growing interest in data center development which reflects the significant economic potential available. States must weigh the costs and benefits of building data center infrastructure for their constituents. Effective planning requires balancing benefits with the environmental, infrastructural, and community impacts of data center expansion. Considerations include projected industry growth, alignment with state economic strategies, partnership models, siting decisions, stewardship of energy and water resources, community engagement practices, and risk management for stalled or incomplete projects. States that evaluate these factors holistically are better positioned to leverage sustainable and equitable growth from data center investment.

On November 6, 2025, we brought together a group of local leaders, subject matter experts, and community members to talk openly about the challenges and opportunities emerging in Georgia as data center development continues across the state. The workshop was a half-day convening that began with quick presentations to provide baseline information on data center development in Georgia, and described water, energy, and community concerns. The group also proposed policy solutions to address these concerns. We provide a summary of this convening and propose a policy roadmap that may be useful for policymakers in Georgia and the rest of the nation.

## II.   Methods

This report draws on insights generated during a facilitated convening held in Atlanta, Georgia, in November 2025. The purpose of the meeting was to examine the policy implications of rapid data center growth in the Atlanta metro region, an area increasingly targeted for large-scale digital infrastructure development. Rising demand for cloud storage, AI, and high-capacity computing has accelerated data center siting across Georgia, reshaping utility planning, environmental policy, and community development. The convening was designed to gather expert perspectives on these trends and to explore potential pathways for equitable and sustainable governance. Additional insights were drawn from a variety of scholarly and media sources covering data centers.

### *A. Structure of the Convening*

The meeting consisted of two components: (1) expert presentations, and (2) facilitated small-group discussions.

We invited five subject-matter experts representing water policy, energy systems, environmental advocacy, regional planning, and state-level politics. Each panelist delivered a brief talk framing one dimension of the data center landscape in Georgia:

- William D. Bryan, Southeast Energy Efficiency Alliance – Energy demand, grid capacity, and decarbonization pathways
- Connie Di Cicco, Georgia Conservation Voters – Community impacts, governance challenges, and potential legislative approaches
- Danny Johnson, Atlanta Regional Commission – Water supply, drought resilience, and regional planning considerations
- Chris Manganiello, Chattahoochee Riverkeeper – Watershed impacts and water-sector regulatory issues
- Amy Sharma, Science for Georgia – Technological trends, siting patterns, and long-term uncertainties

These talks provided participants with a shared foundation of facts, definitions, and emerging concerns.

*B. Breakout Discussions*

Following the presentations, participants were divided into small working groups to examine challenges and opportunities in greater depth. The breakout structure was modeled on "world-building" and future-scenario techniques commonly used in policy design. Participants were asked to imagine the future of data center development in Georgia under different governance models and to identify what safeguards, planning tools, or regulatory frameworks would be necessary to support community well-being while enabling economic growth.

Discussions focused on four themes:

- **Water** – cooling systems, drought conditions, regional water withdrawals, and regulatory gaps.
- **Energy** – grid strain, natural gas and renewable generation, load management, and rate design.
- **Community and Regional Development** – resource utilization and cost, zoning conflicts, equity concerns, employment expectations, and transparency.
- **Policy Pathways** – statewide definitions, siting standards, environmental reporting, and community benefit requirements.

Facilitators captured key points from each group, with particular attention to patterns that emerged across multiple discussions.

*C. Synthesis and Analysis*

The findings presented in this report reflect three sources of analysis:

1. **Participant insights** gathered during the expert presentations and breakout groups.
2. **Cross-theme synthesis** of issues that appeared consistently across water, energy, and community discussions.
3. **A targeted literature review** conducted after the convening to provide national context, refine definitions, and strengthen the introduction and legislative recommendations.

The resulting policy roadmap is grounded in the lived experiences of Georgia communities, the expertise of environmental and energy practitioners, and existing research on data center development in the United States. Its purpose is to inform state and local officials as they consider regulatory frameworks for this rapidly expanding sector.

## III. Findings

The rapid proliferation of data centers across Georgia presents economic opportunities and significant environmental challenges, particularly concerning the state's vital water and energy resources. This discussion aims to provide a comprehensive and localized understanding of these impacts by community experts, offering actionable policy considerations to ensure sustainable growth that benefits all Georgians. As data center development accelerates, their demand for resources is expected to be substantial, ranging from water consumption comparable to entire communities to unprecedented requirements for new generation capacity. Our analysis will emphasize transparency needs, equitable distribution of costs to developers, and the implementation of advanced technologies to mitigate adverse environmental effects.

## A. *Water*

Water is an essential resource for the operation of modern data centers, primarily used for cooling their heat-generating equipment. As data center development in Georgia surges, driven in part by economic development strategies that position Atlanta as a growing tech hub, demand for water is increasing and placing new pressure on the state's finite freshwater supplies. Georgia's hydrogeological context, characterized by reliance on major river basins such as the Chattahoochee River and critical reservoirs like Lake Lanier, heightens the sensitivity of water allocation decisions. The state has a long history of water management challenges, including severe droughts (*Georgia | Drought.Gov*, 2026), as well as ongoing interstate disputes over the Apalachicola-Chattahoochee-Flint (ACF) and Alabama-Coosa-Tallapoosa (ACT) river basins ("Tri-State Water Wars Overview," 2023). These broader constraints frame the long-term availability of water across the region.

In the short term, however, it is municipal water systems that experience the most immediate and tangible impacts of data center growth. While river basins and reservoirs determine overall supply, data centers typically draw water directly from municipal utilities rather than from independent sources. This means that cities and counties must deliver, treat, and distribute large volumes of water using infrastructure originally designed to serve residential and commercial customers. Municipal systems also bear responsibility for wastewater treatment and discharge, further amplifying operational and financial strain. As a result, the effects of increased industrial demand are felt first at the local level, through capacity constraints, infrastructure upgrades, and rising operational costs.

The primary challenge posed by data centers is their substantial water withdrawal, which can strain existing municipal water infrastructure, particularly in smaller communities or areas with aging systems. In Georgia, these facilities are expected to require water equal to as much as 25 percent of a local community's supply (Osaka, 2023). Unlike gradual population growth, data centers can introduce sudden, concentrated demand that exceeds planning assumptions and forces utilities to accelerate capital investments. These pressures may ultimately be passed on to residents and small businesses through higher water rates, even when the benefits of data center development accrue elsewhere.

Traditional open-loop evaporative cooling systems, commonly employed in data centers, lead to significant evaporative water loss, meaning water is consumed rather than returned to the source. This type of consumption can create direct competition with other critical water users, including agriculture, residential communities, and other industries, especially during periods of water scarcity. While less common for cooling discharge, large-scale withdrawals could also indirectly impact water quality by affecting stream flows and concentrating pollutants if not managed carefully (McCauley, 2025). The Georgia Environmental Protection Division (EPD) plays a crucial role in water management, allocation, and permitting, and the increasing demand from data centers adds complexity to balancing competing water needs across the state. In particular, Georgia EPD reports that water resources south of the metropolitan Atlanta region are considered too small to provide substantial amounts of water for municipal and industry supply, yet this area is where new data centers are often sited (Clarke & Peck, 1991).

While water remains a critical resource in Georgia, there are opportunities to mitigate the water challenges posed by data centers through technological advancements and strategic management. Advanced cooling technologies, such as closed-loop systems, significantly reduce water consumption by recirculating cooling water. Other innovations, including direct liquid cooling (which uses a specialized fluid to cool components directly) and adiabatic cooling (which uses less water than traditional evaporative methods), further enhance water (McCauley, 2025). Data centers can also implement on-site water recycling and reuse programs, treating and repurposing their cooling tower blow-down or other process water for non-potable uses. Smart water management systems, incorporating real-time monitoring and demand-side controls, can optimize water use (*Water Metering Resources*, 2026). Furthermore, policies that offer financial incentives or regulatory frameworks encouraging the adoption of these water-efficient technologies can drive widespread implementation, aligning with Georgia's broader water conservation goals and existing watershed management initiatives (Clarke & Peck, 1991).

As discussed in a community workshop led by local Georgia experts, policy options for managing water use at data centers in Georgia could be enhancing water efficiency, ensuring equitable resource distribution, and promoting transparency. The state could consider implementing water-use caps or limits, especially during declared drought conditions, aligning these policies with the Georgia EPD's established drought response plans. Tiered water pricing for large industrial users such as data centers, where higher consumption incurs progressively higher per-unit costs, can incentivize conservation. Furthermore, integrating comprehensive water availability assessments and metrics, such as a resilience score, could be employed to evaluate potential impacts on local water supplies and ecosystems. Additionally, expanding public reporting requirements for data center water consumption is essential for transparency and informed decision-making by local communities and state regulators, ensuring the Georgia Public Service Commission (PSC) can consider these factors when reviewing utility rates. The current information gap and lack of transparency regarding the collective impact of these facilities highlights an urgent need for comprehensive modeling and data collection efforts, potentially involving Georgia's university system, to inform water management strategies. And through entities like a proposed Atlanta Data Center Commission or existing regional water councils, a more holistic approach to water allocation and planning could be achieved.

### *B. Energy*

Data centers are currently known for being energy-intensive facilities that influence global infrastructure and innovation. Their rapid expansion in Georgia, particularly in the Atlanta metropolitan region, is significantly reshaping the state's electricity system. Because data centers operate continuously and require large amounts of power for both computing and cooling, their growth is driving a sharp increase in electricity demand that utilities must plan for years in advance. Meeting this demand is expected to require substantial new generation capacity, new transmission infrastructure, and upgrades to local distribution systems.

Georgia's current energy mix relies primarily on natural gas and nuclear generation, including Plant Vogtle, with a growing but still limited contribution from renewable sources (*Plant Vogtle*, 2026). When large, always-on facilities such as data centers enter the system at scale, they introduce a new class of demand that differs from traditional residential or commercial growth. This demand directly impacts major utilities such as Georgia Power as well as Electric

Membership Corporations (EMCs) across the state, requiring coordinated planning to maintain reliability, affordability, and long-term sustainability.

The first-order effect of data center growth is pressure on transmission and distribution infrastructure. Concentrated, high-load facilities often require new substations, transmission reinforcements, or feeder upgrades. If these investments primarily serve a single industrial customer, they raise difficult questions about cost allocation. Without clear regulatory standards, the costs of infrastructure built to serve data centers may be partially shifted onto residential and small business ratepayers. At the same time, the need to bring new generation online quickly can increase reliance on fossil fuel resources, particularly when renewable generation and storage are not available at sufficient scale. This dynamic risks undermining Georgia's clean energy goals and increasing the state's overall carbon footprint (Sci4Georgia, 2024).

Data centers also affect system reliability through their contribution to peak demand. During hot Georgia summers, cooling systems operate at maximum capacity precisely when residential air-conditioning demand is highest. This overlap intensifies grid stress, increases the risk of service disruptions, and raises system costs during periods of constrained supply. Unlike gradual population growth, data center demand often arrives suddenly and at large scale, compressing planning timelines and amplifying system risk.

Connecting data centers to the grid further illustrates this cause-and-effect chain. Interconnection requires extensive engineering studies to assess feasibility, system impacts, and required upgrades, followed by regulatory approval and construction. These processes can take months or years, delaying project timelines but serving an important function by safeguarding grid reliability. When interconnection queues become long or uncertain, developers increasingly seek alternative power solutions such as on-site natural gas generation, battery storage, or direct power purchase agreements. While these approaches can accelerate deployment, they introduce new regulatory, environmental, and permitting challenges and may further entrench fossil fuel use if not carefully governed.

Workshop participants emphasized that many of these challenges are exacerbated by limited understanding of how data center energy demand varies over time. What participants described as "clean power testing" refers to the lack of detailed load profiles showing when data centers draw the most electricity and whether that demand aligns with renewable generation. Grid planning and stress testing focus on peak demand periods, which often coincide with declining solar output and greater reliance on fossil fuels. As a result, data center approvals are frequently based on projected rather than observed load behavior, creating uncertainty about whether these facilities can operate on clean energy during the hours when they consume the most power. Participants stressed the need for greater research capacity to evaluate load flexibility, storage options, and renewable portfolios before facilities are approved, rather than retrofitting solutions after infrastructure is built.

Regulatory oversight plays a central role in managing these effects. Georgia is not part of a centralized wholesale electricity market; power is exchanged through bilateral arrangements between utilities and generators. As a result, the key policy question is not wholesale price volatility, but how infrastructure costs are evaluated and allocated. Georgia's PSC is responsible for approving utility rate cases and infrastructure investments and must determine whether new facilities built to serve data centers provide system-wide benefits or primarily serve individual large users. These determinations directly affect whether costs are absorbed by data center operators or passed on to other customer classes.

Participants also highlighted inconsistencies in how energy-intensive facilities are regulated across the state. The growth of cryptomining operations in southern Georgia illustrates how large electrical loads can emerge under different regulatory expectations. Unlike data centers, which support a range of digital services, cryptomining facilities serve a narrow private purpose (Science 4 Georgia, 2025). Operating in a relatively nascent regulatory environment, these facilities often face uneven scrutiny regarding energy impacts and cost responsibility (Schmidt, 2025). This patchwork approach complicates statewide planning and undermines consistent energy governance.

Despite these challenges, participants identified significant opportunities to integrate data centers more responsibly into Georgia's energy system. Strategies discussed included requiring renewable energy sourcing, installing solar generation on data center roofs, and parking structures, enforcing efficiency standards such as Energy Star servers, deploying virtual power plants to support load balancing, prioritizing battery-based backup systems over diesel generators, and leveraging community benefit agreements to support residential weatherization and energy efficiency. However, participants cautioned that without clear regulatory incentives and requirements, developers are unlikely to adopt these measures voluntarily.

To manage the energy impacts of data center growth, policymakers could pursue a coordinated regulatory approach focused on transparency, efficiency, and fairness. Requiring disclosure of energy use data and enforcing Power Usage Effectiveness standards would improve planning and accountability. Load-balancing requirements could enable data centers to reduce consumption during periods of grid stress, helping avoid costly new generation. Oversight mechanisms could be established to ensure that infrastructure costs are fairly allocated and that ratepayers are protected from disproportionate impacts. Together, these measures would allow Georgia to support digital infrastructure growth while maintaining grid reliability and protecting communities from unintended energy burdens.

### C. Community & Regional Issues

Community concerns about data centers in Georgia center on the scale of development, the lack of transparency in siting decisions, and the uneven distribution of burdens and benefits. Residents often encounter these facilities not as a single, well-planned project but as clusters of large, resource-intensive buildings concentrated in corridors with existing transmission lines and fiber networks. In many cases, proposed sites were previously agricultural land, forested areas, or low-density neighborhoods. For residents, the visible impacts include noise from cooling equipment and backup generators, light pollution from 24-hour operations, prolonged construction activity, and the replacement of open space with expansive windowless structures.

These concerns are magnified by the absence of a statewide definition or land-use category for data centers. Without consistent guidance, local governments label them in a wide range of categories such as light industrial, heavy industrial, utility, warehousing, or special use. This inconsistency allows large hyperscale facilities to be proposed in areas never designed to absorb high-intensity infrastructure. Participants noted stark differences between the smaller data centers historically found within the city of Atlanta and the massive hyperscale projects now proposed in outer counties. The Quality Technology Services facility on Atlanta's West Midtown/Westside, for example, covers an enormous footprint while employing far fewer people than a city like Atlanta might anticipate for such large facilities, raising questions about whether zoning codes adequately reflect the realities of modern data center operations.

Lack of transparency further fuels community frustration. Residents frequently learn of proposed facilities only when a zoning notice appears or when journalists uncover filings submitted under generic LLC names. Developers have been known to file the same proposal in multiple municipalities simultaneously to identify the fastest approval path or the most generous incentives, reinforcing perceptions that siting decisions prioritize developer convenience over community well-being. Wealthier neighborhoods often have the political resources to organize quickly and block unwanted projects, whereas lower-income and minority communities may find themselves absorbing the cumulative burdens of noise, traffic, and infrastructure expansion with far less influence over the outcome.

Economic concerns add to community skepticism. Data centers occupy large parcels of industrial land yet employ relatively few full-time workers, typically between 30 and 60. This low job-to-land ratio raises questions about whether scarce industrial land is being used efficiently, especially when such land could support hospitals, manufacturing, logistics hubs, or small businesses that generate significantly more employment. At the same time, the infrastructure upgrades needed to serve data centers such as new substations, additional transmission lines, expanded water systems may increase utility costs for residents and small businesses even when the direct benefits, such as tax revenue or "tech hub" branding, flow primarily to state or county governments.

These issues are compounded by a fragmented regulatory environment. Georgia's PCS does not consistently govern how data centers are classified for rate purposes or how utilities design service agreements for them. Local governments lack a standard playbook for negotiating with multibillion-dollar corporations, and many report limited capacity to evaluate long-term water, energy, and infrastructure demands. Without statewide guidance, outcomes vary widely by county and depend more on local political capacity than on any consistent planning standard.

Workshop participants identified several steps that could address these concerns and strengthen community participation. These included requiring developers to begin engagement early in the process, holding public meetings within a defined radius of the proposed site, and following clear, standardized notification timelines. Participants also called for environmental and infrastructure impact reports that are published before any zoning or permitting decision, allowing residents to understand projected water use, energy demand, noise levels, and traffic impacts. Communities expressed strong interest in requiring formal community benefits tailored to project scale, such as electric vehicle charging stations, broadband upgrades, job-training programs, green space, and stormwater improvements. An equity scorecard could help evaluate proposed sites by examining cumulative burdens and historical land-use patterns, ensuring that siting decisions do not disproportionately impact communities with less political power.

By establishing clear expectations for engagement and community benefits, Georgia can give residents a meaningful role in shaping how data centers are built and ensure that the communities hosting these facilities share in their advantages. A more consistent statewide approach would reduce conflict, improve transparency, and support more strategic siting decisions as data center development continues to expand.

### *D. Policy Considerations*

As the need for data infrastructure expands and development continues across Georgia, the state faces an urgent need for clear, forward-looking legislation to strike a balance between innovation and community health. (See https://www.legis.ga.gov/search to track Georgia Legislation.) The following legislative suggestions were proposed by the working group:

**HB 528 – High Resource Use Facilities Transparency Act;**
**Status:** HB 528 remains introduced and pending committee review as of late 2025.
House Bill 528 was filed in the 2025-26 Georgia General Assembly. It proposes to require certain high resource use facilities — defined as those with a peak energy load of at least 30 megawatts — to provide disclosures regarding community impact and energy and water usage. These disclosures would be submitted *before* applying for tax incentives or permits and include information on energy use, water use, and community impacts.

**HB 559 – Data Center Tax Break Sunset Change**
**Status:** HB 559 remains introduced and pending committee review as of late 2025.
House Bill 559 proposes to shorten the sunset on data center tax breaks from 2031 to 2026. While not directly about resource use or community impact, this legislation relates to the broader policy conversation about the *tradeoffs and benefits* data centers receive from the state. Though it doesn't set new community safeguards, changing incentive duration affects the economics of siting decisions.

**SB 34 – Ratepayer Protections Related to Utility Costs**
**Status:** SB 34 remains introduced and pending committee review as of late 2025.
Senate Bill 34 focuses on consumer protections tied to utility cost allocation. According to advocacy summaries, SB 34 seeks to ensure that added infrastructure costs for power upgrades (driven in part by data center load growth) are not paid by Georgia Power customers. While it doesn't itself fix rate design for data centers, it addresses *who pays* for energy infrastructure expansions.

These following policy suggestions were developed from comments made by panelists, participants, legislators, and material provided during the convening. This overview is general and, in some places, vague because it is based on stakeholder comments that were not well described. It is intended to be a high-level plan, capturing the major policy proposals from a variety of engaged stakeholders. The intent is that these ideas lead to a specific plan for achieving a positive impact for the nation, state and communities as data centers expand and infrastructure develops.

**Transparency:** Legislation requiring operational transparency rulings would establish statewide standards that define zoning regulations for data centers, classify such facilities for regulatory and rate purposes, and ensure that communities can benefit from these businesses.

**Zoning:** Legislation to establish clear guidelines for the zoning of data centers, incorporating their staffing needs, power and water usage, and noise impacts on surrounding areas (Hamilton, 2024; Turek & Radgen, 2021).

**Water sources:** Legislation that addresses water use and the stress on systems within operational areas. A current solution to some water demands could be the requirement for closed-loop cooling systems for any new data centers built in Georgia to avoid high evaporative water losses and reduce strain on municipal systems. Additionally, creating limits for data center water usage in times of drought, when water is limited due to the Atlanta metro area's reliance on surface water (*Where Does Atlanta Water Come From?*, 2022) is needed.

**Energy sources:** Legislation mandating renewable energy sources to meet data center energy demands is needed. By developing distributed generation to meet data center demand, these projects will provide a more sustainable resource base and benefit communities that are not shouldering large infrastructure costs for a single user.

**Service Agreement:** Legislation regulating service agreements, allowing data centers flexibility to match power demand to local cycles. By regulating energy draw during high usage times, such as early morning and evenings, when more people are at home, the need for new generation can be limited. By matching service times to low community demand, data centers can reduce energy use for HVAC cooling of interior spaces during high-demand periods on the Georgia grid.

**Cost:** Legislation regarding the cost ratio of these utilities to communities compared to data centers should be a focus to ensure that the subsidization of these facilities does not overburden communities. Legislation limiting the cost these facilities pay to have a maximum ratio below the standard rate paid by community members is needed. Due to their large usage requirements data centers have a negotiating advantage with utilities over the standard ratepayer, legislation could level the playing field between data centers and residential ratepayers.

**Rates:** Local governments need state-level guidance from the Public Service Commission and clear authority to negotiate ordinances and rate structures. With federal initiatives moving to limit state regulatory power, Georgia's leadership has a narrow window to establish a durable framework that protects its autonomy. This guidance would allow municipalities to better negotiate with data center operators developing projects within their communities and to provide a baseline of protection for community members living there.

**Orphaned and abandoned wells:** It is possible that data centers companies will close or relocate. Policies modeled after orphaned oil and natural gas well legislation are needed to protect against a potential collapse. Laws for orphaned and abandoned wells include establishing a bond, trust, or maintenance fund to cover decommissioning, environmental remediation, and infrastructure restoration. Similar to oil wells, these funds could be paid by data center operators during construction and operation. Such measures would ensure that local governments and residents are not left to shoulder the costs of dismantling large-scale facilities, restoring sites, or managing stranded electrical and water infrastructure once operations cease.

## IV.   Conclusion

Georgia stands at an inflection point. The state is one of the top markets for data centers which presents extraordinary opportunities for economic development, technological leadership, and long-term competitiveness in the AI economy. Data Centers support the digital services that households and businesses rely on from sending an email to running an AI model. By shaping how these facilities are planned, sited, and integrated into communities, Georgia can ensure that data centers continue to deliver economic and technological benefits while protecting the resources and neighborhoods that make the state competitive.

Data centers also challenge the state's planning frameworks, utility regulatory structures, and existing approaches to community protection. Data centers are not simply another category of commercial real estate. They are water intensive, energy intensive, and land intensive buildings, and unlike one-off developments such as a water park, golf course or factory, multiple data centers are being proposed in close proximity to one another. This proliferation multiplies their impact and requires policymakers to plan for cumulative, not individual, effects. The policy decisions Georgia makes in the next few years will determine whether the state's AI-era growth is resilient and equitable or costly and contested.

Drawing from the workshop's insights, participants developed a set of policy considerations that offer a practical roadmap for safeguarding Georgia communities as data center development continues.

**The first and most urgent need is transparency among the community, developers, and policymakers.** Without accurate and standardized information about how much water and electricity data centers consume, when they draw from the grid, how their cooling systems function, or what their long-term load projections are, state and local planners cannot act responsibly. Requiring all data centers above a certain size threshold to report annual water use, peak and aggregate power consumption, generator runtime, and projected expansion would give policymakers the facts they need to assess cumulative impacts and coordinate infrastructure investment.

**Georgia could benefit from clearer rules is the timeline and process for community notification**. Data center proposals often surface late in the permitting process and are difficult to discover, leaving residents and local governments little time to understand potential impacts or negotiate appropriate conditions. Georgia can improve transparency by requiring developers to file a formal Notice of Intent to Develop well before any zoning or permit application is submitted. This notice should trigger a public disclosure process that includes posting an environmental and infrastructure impact report detailing projected water use, energy demand, noise levels, traffic, stormwater management, and proposed mitigation strategies. The state can also require developers to hold a series of community meetings within a defined window, such as 60 to 120 days before local hearings, ensuring that residents have adequate time to review findings, ask questions, and provide input. Establishing clear expectations for early engagement not only builds trust, but also helps local governments make informed decisions and reduces conflict later in the process.

Beyond transparency, **Georgia should consider a statewide framework for understanding what a data center is and where it belongs.** At present, data centers fall into a gray zone of land-use categories such as warehouse, industrial site, or utility. This leaves counties and cities to improvise zoning interpretations and negotiate projects individually. A clear state definition paired with guidance for local zoning standards would bring predictability to developers and communities alike. Such guidance should recognize that the highest impacts including significant water withdrawals, heavy energy use, and noise from cooling infrastructure require thoughtful placement in industrial or technology corridors rather than in residential or environmentally sensitive areas. Establishing model ordinances for local governments, especially those with limited planning capacity, would reduce conflict and help align development with long-term community goals.

**Protecting Georgia's water resources** will be essential as data center cooling demands rise alongside population growth and climate-driven drought cycles. Policymakers can mitigate risk by setting expectations for modern cooling technologies such as closed-loop or hybrid systems that reduce evaporative losses. They can also establish drought-stage triggers in which nonessential industrial water uses temporarily pause the amount of water a facility is allowed to withdraw from a water source for a period of time so that residents and critical services always have priority access. Clarifying the rules for water withdrawal permitting, particularly for large cooling loads, will also help communities understand how new facilities affect regional water budgets.

**Energy policy must evolve.** As AI workloads increase, data centers could become one of the fastest growing sources of electricity demand in the state. Georgia will need rules that align industrial growth with grid reliability, consumer rate protection, and clean energy objectives. This may include giving utilities authority to require demand-response participation, setting performance expectations for backup power systems that minimize diesel emissions, and encouraging the development of distributed or renewable generation to meet new industrial loads. Policymakers should also ensure that the costs of new generation, substations, and transmission lines that serve data centers are not shifted onto households and small businesses. Ratepayer protections including transparent cost-allocation methods and enhanced oversight by Georgia's Public Service Commission will be essential to maintaining fairness.

Because incentives for data center development already exist in Georgia, **future incentive structures should ensure that benefits flow to developers and to the communities that host them.** Tying incentives to measurable community benefits such as local workforce training, weatherization programs, rooftop solar expansion, improved broadband access, water system upgrades, beautification projects, or financial contributions to the local businesses and non-profits can transform data centers from isolated industrial sites into genuine economic partners. When negotiated clearly, community benefit agreements provide a foundation for trust and meaningful neighborhood participation in decision-making.

**Georgia should plan proactively for the end of the data center life cycle.** These facilities rely on large-scale infrastructure, and their viability is closely tied to technological trends. A decommissioning and restoration requirement, modeled on the state's approach to orphaned industrial sites, would ensure that data centers contribute to a bond or trust fund throughout their operational life. Should a facility close, relocate, or become obsolete, these funds would cover the costs of dismantling equipment, restoring land, and retiring stranded infrastructure. This protects taxpayers and local governments from liabilities that might otherwise persist for decades.

In addition to insights raised during the convening, the authors note that a growing body of community-focused research and advocacy has emerged to help local governments and residents critically evaluate the return on investment of data center projects. Toolkits developed by organizations such as MediaJustice and Masheika Allgood provide practical guidance for communities seeking to assess proposed developments beyond headline promises of economic growth, including questions about employment, land use, energy demand, water consumption, and long-term public costs. Reports such as *Where the Cloud Meets Cement* by Barakat and colleagues further document how data centers function as permanent industrial infrastructure with localized environmental and fiscal impacts that are often overlooked in early siting decisions. Collectively, these resources emphasize the importance of transparency, early engagement, and rigorous cost-benefit analysis, reinforcing the workshop's conclusion that communities need clearer frameworks and accessible tools to evaluate whether data center projects align with local priorities and deliver durable public value.

Taken together, these steps create a coherent policy roadmap for Georgia. This approach prioritizes transparency, protects water and energy resources, ensures community benefit, and guards against long-term risk. As the state navigates rapid expansion in AI and cloud computing infrastructure, this balanced strategy allows Georgia to capture economic opportunities while safeguarding the well-being of the communities that make that growth possible. Much like the early years of cell tower build-out, Georgia now has the opportunity to establish flexible and future-ready rules that evolve with technology. With proactive planning and collaborative governance, the state can position itself not only as a leader in the digital economy but also as a model for responsible and equitable data center development for the nation.

# V.     References

Clarke, J. S., & Peck, M. F. (1991). *Ground-Water Resources of the South Metropolitan Atlanta Region, Georgia*. USGS.
*Data Center Watch Report Q2 2025*. (2025). Data Center Watch. https://www.datacenterwatch.org/q22025
Erlich, M., & Grabelsky, J. (2005). Standing at a crossroads: The building trades in the twenty-first century. *Labor History*, *46*(4), 421–445.
*Georgia | Drought.gov*. (2026). Drought.Gov. https://www.drought.gov/states/georgia
Greenstein, S., & Fang, T. P. (2022). Where the cloud rests: The location strategies of data centers. *Harvard Business School Working Paper 21-042*.
Hamilton, S. (2024, July 14). *Challenges in Urban Data Center Design* [Text]. Institute of Noise Control Engineering. https://doi.org/10.3397/NC_2024_0089
Ho, A. (2025, November 4). *What you need to know about AI data centers*. Epoch AI. https://epoch.ai/blog/what-you-need-to-know-about-ai-data-centers
Iftekhar, R., & Browne, O. (2025, March 17). *Estimating Data Center Power Demand*. WestWater Research. https://westwaterresearch.com/wmi-estimating-data-center-power-demand/

Jakubiak, J. (2025, October 31). *Data centers as engines of economic growth*. Utilitydive.Com. https://www.utilitydive.com/news/data-centers-as-engines-of-economic-growth/804075/

Leppert, R. (2025, October 24). What we know about energy use at U.S. data centers amid the AI boom. *Pew Research Center*. https://www.pewresearch.org/short-reads/2025/10/24/what-we-know-about-energy-use-at-us-data-centers-amid-the-ai-boom/

Mandler, C. (2024, September 20). Three Mile Island nuclear plant will reopen to power Microsoft data centers. *NPR*. https://www.npr.org/2024/09/20/nx-s1-5120581/three-mile-island-nuclear-power-plant-microsoft-ai

Mayer, V., & Velkova, J. (2023). This site is a dead end? Employment uncertainties and labor in data centers. *The Information Society*, *39*(2), 112–122.

McCauley, P. (2025, August 26). *Data centers consume massive amounts of water – companies rarely tell the public exactly how much*. The Current. https://thecurrentga.org/2025/08/26/data-centers-consume-massive-amounts-of-water-companies-rarely-tell-the-public-exactly-how-much/

Muschter, R. (2025, December 17). *Data center construction in the United States*. Statista. https://www.statista.com/topics/13572/data-center-construction-in-the-united-states/

Osaka, S. (2023, April 25). *A new front in the water wars: Your internet use*. The Washington Post. https://www.washingtonpost.com/climate-environment/2023/04/25/data-centers-drought-water-use/

*Plant Vogtle*. (2026). Georgia Power. https://www.georgiapower.com/about/energy/plants/plant-vogtle.html

Remington, J., & Carter, R. (2025). *An Overview of State Data Center-related Tax Incentives*. Naiop.Org. https://www.naiop.org/research-and-publications/magazine/2024/Winter-2024-2025/development-ownership/an-overview-of-state-data-center-related-tax-incentives/

Robertson, D. (2024, March 22). *The History of Data Centers: An Exponential Evolution*. Enconnex. https://blog.enconnex.com/data-center-history-and-evolution

Schmidt, V. (2025, February 10). *Data Center vs Crypto Mining: Costs & Benefits Compared*. CGAA. https://www.cgaa.org/article/data-center-vs-crypto-mining

Sci4Georgia. (2024). *About Georgia's 2024 Clean Energy Plan*.

Science 4 Georgia. (2025). *Protecting Communities from Crypto Mining Excess – Science for Georgia*. https://scienceforgeorgia.org/knowledge-base1/protecting-communities-from-crypto-mining-excess/

Spindler, W., Hahn-Petersen, L., & Fisher, L. (2025, November 18). *Data centres use vast amounts of water*. World Economic Forum. https://www.weforum.org/stories/2025/11/data-centres-and-water-circularity/

Taylor, P. (2025, November 19). *Data centers worldwide by country 2025*. Statista. https://www.statista.com/statistics/1228433/data-centers-worldwide-by-country/?srsltid=AfmBOoqJPV7WkkPePuvBkzqiba5ZeKp9rQ_nHZXKWEwUhBayjd-oRAWn

Tri-State Water Wars Overview. (2023, November 14). *ARC*. https://atlantaregional.org/what-we-do/natural-resources/tri-state-water-wars-overview/

Turek, D., & Radgen, P. (2021). Optimized data center site selection—Mesoclimatic effects on data center energy consumption and costs. *Energy Efficiency*, *14*(3), 33. https://doi.org/10.1007/s12053-021-09947-y

*USA Data Centers*. (n.d.). DataCenterMap. Retrieved January 22, 2026, from https://www.datacentermap.com/usa/

Walcott, S. M., & Wheeler, J. O. (2001). Atlanta in the telecommunications age: The fiber-optic information network. *Urban Geography*, *22*(4), 316–339.

*Water Metering Resources*. (2026). Energy.Gov. https://www.energy.gov/femp/water-metering-resources

*Where does Atlanta water come from? | Southern Green Inc.* (2022, March 2). https://www.southerngreen.com/blog/where-does-atlanta-water-come-from-and-where-does-the-wastewater-go